\documentclass[pre,aps,twocolumn]{revtex4} 
\usepackage{graphicx} 
\usepackage[usenames]{color}
\begin{document} 

\title{A microscopic view on contact angle selection}

\author{Jacco H. Snoeijer} 
\affiliation{School of Mathematics, University of Bristol, University Walk, Bristol BS8 1TW, United Kingdom}

\author{Bruno Andreotti}
\affiliation{Physique et M\'ecanique des Milieux H\'et\'erog\`enes, UMR
7636 ESPCI -CNRS, Univ. Paris-Diderot, 10 rue Vauquelin, 75005, Paris,
France}
 
\date{\today} 

\begin{abstract}
We discuss the equilibrium condition for a liquid that partially wets a solid on the level of intermolecular forces. Using a mean field continuum description, we generalize the capillary pressure from variation of the free energy and show at what length scale the equilibrium contact angle is selected. After recovering Young's law for homogeneous substrates, it is shown how hysteresis of the contact angle can be incorporated in a self-consistent fashion. In all cases the liquid-vapor interface takes a nontrivial shape, which is compared to models using a disjoining pressure. 
\end{abstract} 

\maketitle 

\section{Introduction}
The equilibrium condition of a liquid that partially wets an homogeneous solid substrate has been addressed since Young \cite{Y1805}, who found that there is a well-defined contact angle $\theta_Y$. This angle minimizes the macroscopic free energy $E$ of the liquid and provides the boundary condition for the free surface: 
\begin{equation}\label{young}
\cos \theta_Y = \frac{\gamma_{sv}-\gamma_{sl}}{\gamma},
\end{equation}
where $\gamma$, $\gamma_{sl}$ and $\gamma_{sv}$ represent the surface tensions of the liquid-vapor, solid-liquid and solid-vapor interfaces respectively. The surface tensions are defined as the excess free energy per unit area, in the particular geometry of a planar interface between two unbounded phases~\cite{KB49,RW82}. When approaching the contact line where the three interfaces meet, the geometry changes dramatically and the force balance can no longer be expressed in terms of surface tensions. Instead, the interface deforms at small scales to establish a nontrivial equilibrium shape~\cite{PGG85,MK92}. Macrocopically, however, one recovers a wedge of angle $\theta_Y$ \cite{CW91}. 

The small scale structure of the contact line is much less understood, but very relevant for problems such as line tension in nanofluidics~\cite{DI93,GD98,PH00,HPF00}, or moving contact lines that are out-of-equilibrium down to molecular scales~\cite{HS71,V76,PP00,E04,DFSA08,BEIMR08}. To avoid fitting parameters, these problems require an explicit treatment of the force balance within the range of molecular interactions. Another basic phenomenon not described by Young's law is contact angle hysteresis: chemical inhomogeneities or roughness of the solid substrate can trap the contact line in a potential well, allowing for a range of possible macroscopic contact angles~\cite{DC83,PGG85,RG07}. A common interpretation is to assume that the contact angle is selected at a scale smaller than the inhomogeneities, and to consider Young's law as a local boundary condition~\cite{Joanny,SG86,NB03}. Strictly speaking, however, this interpretation is not self-consistent because surface tensions are no longer well-defined at molecular scales. Alternatively, one could argue that the global free energy $E$ is dominated by contributions from the bulk of the drop, and described by an average over the inhomogeneities, $\bar{\gamma}_{sv}-\bar{\gamma}_{sl}$~\cite{W36,CB44}. While this approach has been very successful in recent years for wetting on textured substrates~\cite{BTQ01}, it predicts a unique value for the contact angle and hence does not capture the hysteresis. 

These observations raise the question of how the contact angle is selected, in particular at what length scale. Rather than imposing the angle as an external boundary condition at a large distance from the contact line~\cite{PGG90}, one would like to {\em derive} it directly from the force balance at molecular level.
\begin{figure}[t!]
\includegraphics{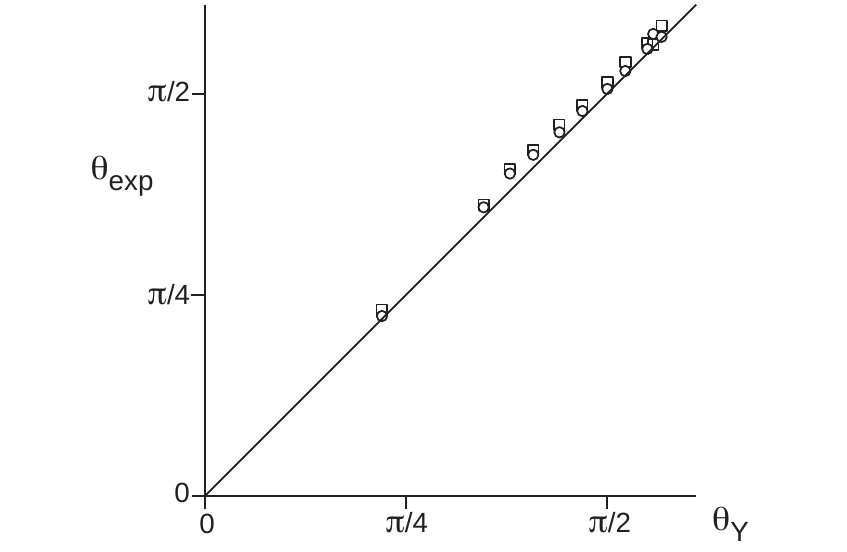}
\caption{Experimental verification of Young's law by~\cite{CW91}: the measured contact angle $\theta_{\rm exp}$ versus the prediction $\theta_Y$ from independent measurements of the surface tensions.} 
\label{fig.YoungExp} 
\end{figure}

There are various approaches to address this problem within a continuum theory, but no general consensus has yet been achieved. A popular strategy consists of adding the microscopic effect of liquid film confinement (disjoining pressure) and the macroscopic effect of interface curvature (Laplace pressure)~\cite{PP00,DI93,S98}. A consequence of this approximation is that Young's angle is only recovered if a precursor film is present on the substrate. Conceptually, both the disjoining and Laplace pressures have the same physical origin: the force on the interface arises because the geometry of the liquid deviates from a semi-infinite flat domain. {\it A priori}, these pressures need not be additive and a more rigorous approach consists of minimizing the total free energy of the system with respect to the interface shape $h(x)$. From this, Getta and Dietrich~\cite{GD98} derived a nonlocal integral equation for $h(x)$ within Density Functional Theory in the presence of a mesoscopic precursor film. They demonstrated that the addition of Laplace and disjoining pressures is in fact a local approximation and they addressed the quantitative limitations of such a local approach. 

The situation without precursor film, typical of more hydrophobic substrates~\cite{BB99,DL06}, has been analyzed by a number of papers~\cite{MK92,KM91,H93,WW04}. Surprisingly, these arrived at contradicting conclusions. Merchant and Keller~\cite{MK92} derived a nonlocal integral equation for the interface profile that has essentially the same structure as in~\cite{GD98}. From asymptotic analysis they showed that the profile macroscopically approaches Young's angle, while the inner structure of the contact line displayed nontrivial oscillations. On the other hand, Hocking~\cite{H93} and more recently Wu and Wong~\cite{WW04} have identified an exact solution of a perfectly straight wedge down to molecular scale. Although unnoticed by these authors, the corresponding wedge angle is not Young's angle as obtained by Merchant and Keller, which raises an intriguing paradox. 

In this paper we revisit the equilibrium condition for the liquid-vapor interface at small length scales. We aim (i) to explicitly identify the connection between microscopic and macroscopic length scales, and thus the selection of the contact angle, (ii) to apply this to contact angle hysteresis, (iii) to clarify some of the mentioned contradictions in the literature. In view of the latter, the descriptions in the paper will in places be somewhat extensive.

The analysis is restricted to mechanical equilibrium and treats the interface as infinitely thin, hence ignoring thermal fluctuations and the corresponding details of the density profile. These simplifications allow writing down a free energy functional $E[h]$ that contains all length scales but yet remains accessible for analysis. The capillary pressure $\Pi$ is generalized as the functional derivative of the free energy with respect to the interface shape $h$. $\Pi$ turns out to be exactly equal to the potential energy associated to the intermolecular forces. The equilibrium condition is then simply a constant capillary pressure, i.e. an iso-potential free surface: $\Pi=\lambda$.

A first result is that the shape of the interface is never a perfectly straight wedge, due to regularization of the van der Waals interactions at a distance $\ell$, typically several Angstroms~\cite{I92}. This regularization is crucial, as otherwise the total energy diverges and surface tensions are not defined. For homogeneous substrates it is found that the solution asymptotically approaches Young's contact angle in the large scale limit, effectively reproducing the result by Merchant and Keller~\cite{MK92}. The small scale structure, however, does not exhibit the oscillations seen in~\cite{MK92}. We then identify how the large scale angle is selected in the neighbourhood of the contact line, at the scale $\ell$ of the regularization, and provide a self-consistent description for contact angle hysteresis. Finally, we compare the generalized capillary pressure to the usual disjoining pressure.

\section{Macroscopic analysis}\label{sec.macro}
The benchmark for our analysis is the usual macroscopic theory that we wish to recover from the intermolecular forces. Here we present a formal derivation (similar to~\cite{BSW95}), which can later on be generalized to include microscopic interactions. We consider profiles that are translationally invariant in one direction, so that the problem reduces to finding $h(x)$. The free energy of the liquid film reads (per unit length $y$),
\begin{eqnarray}
E[h] &=& \int_a^b dx \,\Gamma_{\rm wet} + 
\int_{-\infty}^{a} dx \,\Gamma_{\rm dry} +
\int_{b}^{\infty} dx\, \Gamma_{\rm dry}    \nonumber \\
&=&  \int_a^b dx \,\Gamma(h,h') + const,
\end{eqnarray}
with 
\begin{equation}\label{gammamacro}
\Gamma(h,h') = \gamma \sqrt{1+h'^2} + \gamma_{sl} - \gamma_{sv}.
\end{equation}
where from now on we write $\gamma_{lv}=\gamma$. The factor $\sqrt{1+h'^2}$ is required to compute the surface area (arclength) of the liquid-vapor interface. To minimize this energy under the constraint of a fixed liquid volume one needs to introduce a Lagrange multiplier $\lambda$, and then consider the variation of ${\cal E}=E - \lambda \int dx\,h$, 
\begin{eqnarray}\label{varmacro}
\delta {\cal E} &=& {\cal E}[h+\delta h] - {\cal E}[h]  \nonumber \\
&=& \Gamma(b) \delta b -\Gamma(a) \delta a + \frac{\partial \Gamma}{\partial h'} \delta h |_a^b 
\nonumber \\
&&
+ \int_a^b dx \, \left( 
\frac{\partial \Gamma}{\partial h} - \frac{d}{dx} \frac{\partial \Gamma}{\partial h'} - \lambda \right)\delta h.
\end{eqnarray}
Besides the usual partial integrations, we find contributions $\delta a$ and $\delta b$ that correspond to contact line variations. These arise because there is no external force to constrain the contact line positions, so one can not impose $\delta h=0$ at the boundary.

At equilibrium this variation $\delta {\cal E}=0$ for arbitrary $\delta h(x)$. In the bulk of the liquid, the vanishing of the integral yields the Euler-Lagrange equation:
\begin{equation}\label{eulerlagrange}
\Pi \equiv \frac{\delta E}{\delta h} =\frac{\partial \Gamma}{\partial h} - \frac{d}{dx} \frac{\partial \Gamma}{\partial h'} = \lambda,
\end{equation}
which for $\Gamma(h,h')$ of (\ref{gammamacro}) reduces to the usual Laplace pressure condition
\begin{equation}
\Pi=-\gamma \kappa=\lambda,
\end{equation}
where $\kappa = h''/(1+h'^2)^{3/2}$ is the curvature of the interface.

Similarly to the Hamiltonian in classical mechanics, this differential equation has a first integral, 
\begin{equation}\label{imacro}
{\mathcal G}_{\rm macro} = \Gamma - h' \frac{\partial \Gamma}{\partial h'} - \lambda h
\end{equation}
that is conserved along the free surface. This can be seen by multiplying (\ref{eulerlagrange}) by $h'$ and integrating along $x$. To close the problem, we need to determine the value of ${\mathcal G}_{\rm macro}$ from the contributions at the contact line $x=a,b$ in (\ref{varmacro}). The variation of the contact line $\delta a$ is not independent of $\delta h(a)$. Through Taylor expansion one finds the relation $\delta a=-\delta h(a)/h'(a)$. As the contributions should vanish for arbitrary $\delta h$, this provides so-called natural boundary conditions:
\begin{equation}
-\frac{\Gamma}{h'} + \frac{\partial \Gamma}{\partial h'}=0 \quad {\rm at} \quad x=a, x=b.
\end{equation}
Comparing to (\ref{imacro}), one recognizes that these boundary conditions simply yield ${\mathcal G}_{\rm macro}=0$. For the problem at hand we thus find
\begin{equation}
{\mathcal G}_{\rm macro} = \frac{\gamma}{\sqrt{1+h'^2}} - (\gamma_{sv}-\gamma_{sl}) - \lambda h= 0.
\end{equation}
At the contact line this indeed gives Young's law (\ref{young}).

\section{Generalized equilibrium conditions}
\subsection{Formulation}
Before analyzing the microscopic free energy, let us reiterate the structure of the analysis. The bulk equilibrium condition is obtained from the functional derivative, which basically corresponds to a constant pressure. We therefore introduce the generalized capillary pressure $\Pi$ as the functional derivative, yielding the equilibrium condition in the bulk:
\begin{equation}\label{pressure}
\Pi \equiv \frac{\delta E}{\delta h} = \lambda,
\end{equation}
where $\lambda$ is, again, the Lagrange multiplier associated to incompressibility.

The generalized invariant ${\mathcal G}$ can be constructed from integration
\begin{equation}\label{defi}
{\mathcal G} \equiv \int_a^{\tilde{x}} dx \,h'(x) \left[  \Pi(x) - \lambda \right],
\end{equation}
which from (\ref{pressure}) vanishes for all values of $\tilde{x}$. In the macroscopic calculation we closed the problem from the variation of the contact line, which gave a term $\Gamma(a)\delta a$. This contribution is due to the discontinuous jump from $\Gamma_{\rm wet}$ to $\Gamma_{\rm dry}$ across the contact line. We anticipate that on a molecular level such a discontinuity does not occur, as the relevant energies vary smoothly with the thickness $h$. This implies that in the microscopic model we lose the boundary condition. Instead, the solution will be selected internally from the balance of intermolecular forces. 

\subsection{Interpretation}
The capillary pressure can be interpreted as a purely geometric effect. For a perfectly flat interface between two semi-infinite phases it is zero, because a virtual displacement of the interface will still yield a surface between two unbounded phases. However, any deviation from this geometry will result into a force on the interface. The best known example is of course the Laplace pressure for curved interfaces, for which a virtual displacement leads to a change in surface area. Another example is the disjoining pressure for a thin horizontal liquid film on a solid substrate, whose thickness falls within the range of intermolecular forces. In this case the molecules near the surface feel a change of environment, due to a replacement of liquid for solid molecules. We emphasise that both pressures have the same physical origin and are captured within the single definition (\ref{pressure}). 

The integral ${\mathcal G}$ represents the horizontal component of the total force acting on the liquid between the contact line and the location $x$. Namely, the horizontal force per unit length $y$ acting on a vertical slice of liquid reads:
\begin{equation}
d^2f = - \Pi'(x) h dx\,dy
\end{equation}
The horizontal component of the force is then obtained from integration by parts
\begin{eqnarray}
\frac{df}{dy}&=& - \int_a^x d\tilde{x}\,\Pi'(\tilde{x}) h(\tilde{x}) \nonumber \\
&=& h(a) \Pi(a)- h(x) \Pi(x) +  \int_a^x d\tilde{x} \,h'(\tilde{x}) \Pi(\tilde{x}) \nonumber \\
&=& {\mathcal G}.   
\end{eqnarray}
As we have seen from the macroscopic analysis, this horizontal force balance should yield Young's law for homogeneous substrates. In the presence of contact angle hysteresis, however, the invariant ${\mathcal G}$ provides a nontrivial communication across length scales. We show below how it translates the microscopic force balance into a macroscopic contact angle.

\subsection{Microscopic free energy and capillary pressure}
We consider pairwise molecular interactions, $\phi_{\alpha \beta}(|{\bf r}'-{\bf r}|)$, where $\alpha$ and $\beta$ can represent molecules in the liquid (l) or solid phase (s). We then follow~\cite{KM91}, by considering density-density correlations 
\begin{eqnarray}
\rho_{\alpha \beta}^{(2)}({\bf r},{\bf r}') 
\approx \rho_\alpha \rho_\beta \,g_{\alpha \beta}(|{\bf r}'-{\bf r}|),
\end{eqnarray} 
where $g_{\alpha \beta}$ is the pair correlation function and we take $\rho_{l,s}$ constant over the liquid domain (${\cal L}$) and solid domain (${\cal S}$) respectively. This is equivalent to the so-called sharp-kink approximation in Density Functional Theory \cite{GD98}. In the absence of external forces, this then yields the energy functional

\begin{eqnarray}\label{emicro}
E[h] &=& \frac{1}{2} \int_{\cal L}d{\bf r} \int_{\cal L}d{\bf r}' \,  \tilde{\phi}_{ll}(|{\bf r}'-{\bf r}|) 
\nonumber \\
&&
-  \frac{1}{2}  \int_{\cal L}d{\bf r} \int_\infty d{\bf r}' \, \tilde{\phi}_{ll}(|{\bf r}'-{\bf r}|) 
\nonumber \\
&& + \int_{\cal S}d{\bf r} \int_{\cal L}d{\bf r}' \,  \tilde{\phi}_{sl}(|{\bf r}'-{\bf r}|),
\end{eqnarray}
where we renamed $\tilde{\phi}_{\alpha \beta}(r)=g_{\alpha \beta}(r)\phi_{\alpha\beta}(r)$. As this represents the excess free energy, we subtracted the bulk liquid energy emerging from an infinite domain of interaction. The factors $1/2$ arise since all pairwise integrations are counted twice by the double integrations. By definition, the gaseous domain is the complementary of the solid and liquid domains. One thus obtains an equivalent system if the gas is replaced by vacuum and if the gas-liquid interaction potential is subtracted from the liquid-liquid and the solid-liquid potentials. So, once the potentials are expressed in terms of surface tensions, one rigorously finds the same result if the gas phase is virtually replaced by a vacuum.

We again consider profiles that are invariant along the $y$ direction, bounded by contact lines at $x=a$ and $x=b$. The domains of integration then become
\begin{eqnarray}
\int_{\cal L}d{\bf r}  &=& 
\int_a^b dx \int_{-\infty}^\infty dy \int_0^{h(x)}dz 
\nonumber \\
\int_{\cal S}d{\bf r}  &=&
\int_{-\infty}^\infty dx \int_{-\infty}^\infty dy \int_{-\infty}^0 dz 
\nonumber \\
\int_\infty d{\bf r} &=&
\int_{-\infty}^\infty dx\int_{-\infty}^\infty dy\int_{-\infty}^\infty dz.
\end{eqnarray}
The interface profile $h(x)$ now appears explicitly as the boundary of the liquid domain. This makes it easy to evaluate the functional derivative, i.e. the capillary pressure, as
\begin{eqnarray}\label{pcap}
\Pi(x) &=& \int_a^b dx' \int_0^{h(x')} dz' \,\varphi_{ll}\left(x'-x,z'-h(x) \right)
\nonumber \\
&& - \int_{-\infty}^\infty dx' \int_0^\infty dz'\, \varphi_{ll}\left(x',z'\right)  
\nonumber \\
&&+ \int_{-\infty}^\infty dx' \int_{-\infty}^0 dz' \, \varphi_{sl}\left(x'-x,z'-h(x)\right).  
\nonumber \\ 
&&
\end{eqnarray}
Here we conveniently integrated out the invariant $y$ direction, so that 
\begin{equation}
\varphi(x,z) = \int_{-\infty}^\infty dy \, \tilde{\phi}\left( \sqrt{x^2+y^2+z^2}\right).
\end{equation}

The pressure at the interface can thus be split into a part due to liquid-liquid interaction and a part due to solid-liquid interaction, i.e. $\Pi = \Pi_{ll}+ \Pi_{sl}$. Inspection of the integrals reveals that these are simply the potential energy per unit volume at the free surface, due to the presence of liquid and solid molecules. The equilibrium condition is thus that $h(x)$ is an equipotential~\cite{MK92,H93}. Mechanically, a gradient in potential energy would lead to fluid motion. Note that the recent paper~\cite{WW04} uses the potential on the free surface to estimate the disjoining pressure, but then resides to a local approximation for the functional derivative, similar to equation (\ref{eulerlagrange}).

\subsection{Surface tensions}\label{sec.surfacetension}
Once the molecular interactions have been specified, one can compute the surface tensions~\cite{KB49,RW82}. This is important in order to establish a connection with the macroscopic limit. An elegant way to obtain the liquid-vapor tension $\gamma$ is to directly derive the Laplace pressure from (\ref{pcap}) in the macroscopic limit. In this case the boundary $z=0$ can effectively be replaced by $-\infty$, while $a,b=\pm \infty$. A Taylor expansion $h(x')-h(x)=h'(x)u+ h''(x)u^2/2$ and changing variables $u=x'-x$ then yields~\cite{KM91,GD98}
\begin{equation}
\Pi(x) \simeq - \gamma \kappa,
\end{equation}
with
\begin{equation}\label{gamma}
\gamma =  - \frac{\pi}{2}\int_0^\infty dr \, r^3 \tilde{\phi}_{ll}(r).
\end{equation}
Indeed, this is precisely twice the energy required to separate two semi-infinite liquid domains from contact to infinity. 

The other surface tensions can be derived from the disjoining pressure of a perfectly flat horizontal film, $\Pi_{\rm disj}(h)$, through the connection~\cite{C95,S98}
\begin{eqnarray}
\gamma + \gamma_{sl} - \gamma_{sv} &=& \int_0^\infty dh\, \Pi_{\rm disj}(h).
\end{eqnarray}
This relation expresses that $\Pi_{\rm disj}(h)$ is the derivative of a surface free energy that has the correct macroscopic limits $\Gamma_{\rm wet}$ and $\Gamma_{\rm dry}$. Taking $h(x)=h(x')=h$ and $a,b=\pm \infty$, we find upon integration
\begin{eqnarray}\label{spreading}
\gamma_{sl} - \gamma_{sv} -\gamma &=& \pi \int_0^\infty dr \, r^3 \tilde{\phi}_{sl}(r).
\end{eqnarray}
Note that (\ref{gamma}) is a special case of this result, where the solid is replaced by liquid so that $\gamma_{sv}=0$ and $\gamma_{sv}=\gamma$. For a more rigorous thermodynamic treatment of surface tension we refer to~\cite{RW82}.

\section{Homogeneous substrates}

\subsection{Asymptotic analysis: Young's law}

We now compute the equilibrium contact angle from the microscopic interactions, in the case where the solid surface is perfectly homogeneous. In the macroscopic analysis of Sec.~\ref{sec.macro}, Young's law arises from variation of the contact line position, $\delta a$. If we consider such a variation for the free energy (\ref{emicro}), we find a contribution proportional to $\int_0^{h(a)}dz$. Since $h(a)=0$ by construction, this variation does not give a nontrivial boundary condition. As the contact angle should still emerge from the horizontal force balance, we evaluate the invariant ${\mathcal G}$ by integration (\ref{defi}). If the solid substrates are spatially homogeneous, the solid-liquid potential is a function of the film thickness only:
\begin{eqnarray}
\Pi_{sl}(h) &=& \int_{-\infty}^\infty dx' \int_{-\infty}^0 dz' \, \varphi_{sl}\left(x',z'-h\right).
\end{eqnarray}
It contributes to the invariant as
\begin{equation}
{\mathcal G}_{sl}(x) = \int_0^{h(x)} d\tilde{h} \, \Pi_{sl}(\tilde{h}).
\end{equation}
Integrating to a large distance from the contact line, $h \gg \ell$, this can be further reduced to
\begin{eqnarray}
{\mathcal G}_{sl}(\infty) &=& \pi \int_0^\infty dr \, r^3 \tilde{\phi}_{sl}(r)
\nonumber \\ 
&=&
\gamma_{sl} - \gamma_{sv} -\gamma,
\end{eqnarray}
where we used the definition of surface tensions (\ref{spreading}) of the previous paragraph.

The contribution of the liquid-liquid interaction is more subtle, as it involves the (unknown) equilibrium shape $h(x)$,  
\begin{eqnarray}\label{gll}
{\mathcal G}_{ll}(\infty) &=& \int_0^{\infty} dx  \, h'(x)
\nonumber \\
&&
\left[ \int_0^\infty dx' \int_0^{h(x')} dz' \,\varphi_{ll}\left(x'-x,z'-h(x) \right) \right.
\nonumber \\
&& \left. 
- \int_{-\infty}^\infty dx' \int_0^\infty dz'\, \varphi_{ll}\left(x',z'\right) \right], 
\end{eqnarray}
where we took $a=0$ and $b=\infty$. However, it was shown by Merchant and Keller~\cite{MK92} that the integral can be evaluated analytically if one assumes that for large $h$ the interface approaches a wedge of a well-defined slope, $h(x)\simeq x \tan \theta_\infty$. The trick is to consider the difference between the real $h(x)$ and the straight wedge. Merchant and Keller then showed that this difference vanishes due to the double integration over $x$ and $x'$, irrespective of the shape $h(x)$. In Appendix~\ref{app.MK} we clarify the crucial steps in their analysis. The end result depends only on the value of the asymptotic slope: 
\begin{equation}\label{gllbis}
{\mathcal G}_{ll}(\infty) = \gamma + \gamma \cos \theta_\infty.
\end{equation}
Taking $\lambda=0$, the invariance ${\mathcal G}={\mathcal G}_{ll}+{\mathcal G}_{sl}=0$ yields
\begin{equation}
\gamma \cos \theta_\infty + \gamma_{sl} - \gamma_{sv} = 0.
\end{equation}
Comparing to (\ref{young}), we indeed find $\theta_\infty = \theta_Y$ from Young's argument. So whatever the shape $h(x)$, the equilibrium solution approaches Young's angle for $h\gg \ell$. For finite drop volumes of typical size $R$, corrections are of order $\lambda h \propto \gamma h/R$, so that the asymptotic analysis is justified for $\ell \ll h \ll R$.

To recapitulate, the variation of the contact line position does not provide a boundary condition in the microscopic theory. Instead, there is an internal selection of the solution for which one can derive the asymptotic angle $\theta_\infty=\theta_Y$. This is a completely microscopic demonstration of Young's law. Imposing a wedge angle different from $\theta_Y$, hence ${\mathcal G}\neq 0$, will force the solution to pass through a minimum or a singularity, as only the solution with ${\mathcal G}=0$ can reach $h=0$.

\subsection{Equilibrium profiles}\label{sec.numerics}

To analyze the inner structure of the contact line, we numerically solve the integral equation $\Pi=\lambda$, with the pressure taken from (\ref{pcap}). We consider the following effective interaction
\begin{eqnarray}\label{interaction}
\tilde{\phi}_{\alpha \beta}(r) =
\left\{ 
\begin{array}{l l}
-c_{\alpha \beta}/r^6 & \quad \textrm{for $r \geq \ell$} \\
-c_{\alpha \beta}/\ell^6 &  \quad \textrm{for $r < \ell$}.
\end{array}
\right.
\end{eqnarray}
This corresponds to attractive Van der Waals interactions, regularized below $r = \ell$, combined with a flat pair correlation function,  $g(r)=1$. For simplicity we take the same regularization length $\ell$ for the liquid-liquid and solid-liquid interactions. Using Sec.~\ref{sec.surfacetension}, we can readily compute Young's angle as~\cite{BB99}:
\begin{equation}\label{youngmicro}
1-\cos \theta_Y = 2\left(1- \frac{c_{sl}}{c_{ll}}\right),
\end{equation}
For details on the numerical algorithm we refer to Appendix~\ref{app.numerics}.

\begin{figure}[t!]
\includegraphics{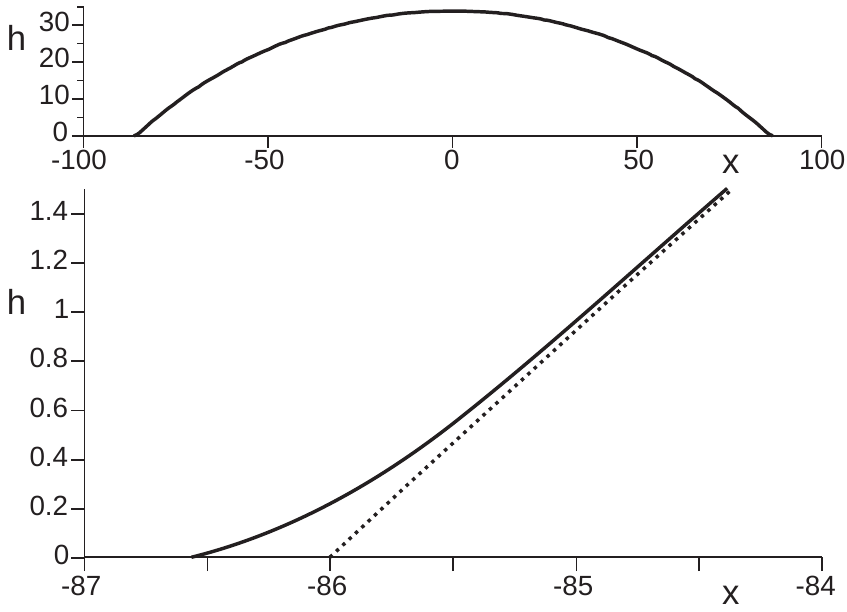}
\caption{Equilibrium profile of a drop with interaction (\ref{interaction}), with $\theta_Y=0.7$. Lengths are expressed in terms of $\ell$. Top: at large scales, the shape is a cylindrical cap. Bottom: a zoom at the contact line region reveals a variation of the slope, from a micrscopic angle $\theta_\mu$ to the a macroscopic angle $\theta_\infty$. The dotted line confirms that $\theta_\infty=\theta_Y$.} 
\label{fig.NanoDrops} 
\end{figure}

Figure~\ref{fig.NanoDrops} shows an equilibrium drop profile, with $\theta_Y=0.7$. Close to the contact line we observe a change of contact angle from a microscopic value, $\theta_\mu$, to the equilibrium angle, $\theta_\infty=\theta_Y$. Indeed, the dotted line indicates Young's angle computed from (\ref{youngmicro}), confirming the asymptotic analysis. This equilibrium angle is attained at a typical height $\ell$ (set to unity in all figures) above the solid. At larger distances, outside the range of the interactions, the profile is simply the cylindrical cap expected from macroscopic theory. 

\begin{figure}[t!]
\includegraphics{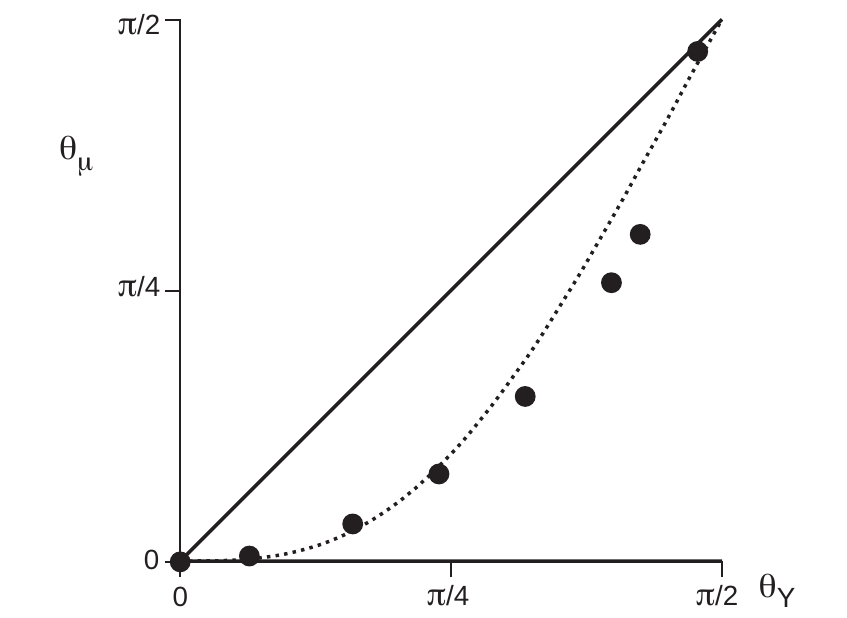}
\caption{Relation between microscopic angle $\theta_\mu$ and Young's angle $\theta_Y$. Dots: numerical resolution for (\ref{interaction}). Dotted line: interpolation based on local approximations (\ref{thetamu1}). Solid line: $\theta_\mu=\theta_Y$, plotted for clarity only. Note that $\theta_\mu=\theta_Y$ only at $\pi/2$.} 
\label{fig.ThetaMicro} 
\end{figure}

In Fig.~\ref{fig.ThetaMicro} we present the microscopic angle as obtained for various wetting conditions. The microscopic and macroscopic angles coincide only for $\theta_Y=\pi/2$, for which the solution is a perfectly straight wedge. For more hydrophilic drops, however, the microscopic angle is always much smaller than Young's angle. In Appendix~\ref{app.local} we derive an approximate relation based on local theories:

\begin{equation}\label{thetamu1}
\theta_\mu \approx  \frac{1}{2} \theta_Y \left( 2-\cos \theta_Y  - \cos^2 \theta_Y\right),
\end{equation}
see also the Discussion section. This provides a good qualitative description of the numerical results (dotted line, Fig.~\ref{fig.ThetaMicro}).

Our numerical results gave no indication of the surprising oscillatory behavior of $h(x)$, as observed by Merchant and Keller~\cite{MK92}. In their simulations, the macroscopic wedge angle was imposed as a boundary condition at infinity, which is different from our finite volume nanodroplets. Despite this technical difference, the inner structure of the contact line should be the same. Investigating the microscopic parameters in~\cite{MK92} it seems that the asymptotic boundary condition was not correctly taken at Young's angle~\cite{footMK}. This means that macroscopically ${\mathcal G}\neq 0$ in their simulations. Enforcing a solution that at the same time has $h=0$, where ${\mathcal G}= 0$, the solution would have to pass through a singularity. Looking at the cusp-like structures in~\cite{MK92}, we think this is a plausible explanation. This problem could not possibly occur in our simulations, as the drop was free to establish the equilibrium shape. 

\subsection{Paradox of the straight wedge}

The findings discussed above contradict the result by Hocking~\cite{H93}, who analytically found equilibrium solutions that remain a perfectly straight wedge at all scales. Hocking's solutions exist for all wetting angles, while in our simulations the wedge was only found for $\theta_Y=\pi/2$. The paper~\cite{H93} starts with the same expression for the capillary pressure (\ref{pcap}), and then uses unregularized Van der Waals interactions 
\begin{eqnarray}\label{hockingphi}
\phi_{\alpha \beta}(r) &=& \frac{- c_{\alpha\beta}}{r^6} \quad \textrm{for $r \geq 0$}.
\end{eqnarray}
Imposing the shape $h(x)=x \tan \theta$ this yields 
\begin{eqnarray}\label{piwedge}
\Pi_{\rm wedge} &=& \frac{\pi c_{ll}}{6h^3} \left( 1 -   \frac{c_{sl}}{c_{ll}} - F(\theta)     \right)
\end{eqnarray}
with $F(\theta) = \frac{1}{2} - \frac{3}{4} \cos \theta  + \frac{1}{4} \cos^3 \theta$. In order for the pressure to be constant along the interface, for all $h$, it is then argued that the equilibrium angle follows from
\begin{equation}\label{thetaH}
F(\theta_H) = 1- \frac{c_{sl}}{c_{ll}} \quad \Rightarrow \quad 
\theta_H \simeq  \left( 1-\frac{c_{sl}}{c_{ll}} \right)^{1/4},
\end{equation} 
where the latter approximation holds for small angles. 

Although the paper claims to be consistent with Young's law and~\cite{MK92}, this is not the case. Comparing (\ref{thetaH}) to (\ref{youngmicro}), it is clear that  $\theta_H \neq \theta_Y$. Crucially, the interactions (\ref{hockingphi}) lack a small-scale regularization and diverge when $r \rightarrow 0$. Integrating over this singularity gives a divergent contribution, so that the surface tensions are effectively infinite. As we have seen, the wedge solutions cease to exist when a regularization is introduced. 

The expression (\ref{piwedge}) nevertheless has a useful interpretation as the scaling for Van der Waals interactions when $h\gg \ell$. In fact, it has been used to derive the approximation (\ref{thetamu1}). For small contact angles, the contribution $F(\theta) \propto \theta^4$ can be neglected so that the disjoining pressure is unaffected by the small inclination. For $\theta_Y = \pi/2$, however, the term in brackets vanishes and the effect of Van der Waals forces completely disappears. In this particular case, the wedge indeed is an exact solution as observed in our simulations. 

\section{Contact angle hysteresis}\label{sec.hysteresis}

\subsection{Analysis}

The microscopic determination of $\theta_\infty$ follows from asymptotic analysis and does not depend on the homogeneity of the solid surface. One can distinguish two types of inhomogeneity that both lead to contact angle hysteresis: geometrical roughness of the substrate and variations in surface chemistry. In the latter case, the analysis of the ${\mathcal G}$ integrals can be done explicitly. Let us therefore consider a spatially varying surface potential,
\begin{equation}
\Pi_{sl} = -\gamma (1+\cos\theta_Y) \Psi^0(h) - \epsilon \gamma \cos(qx+\alpha) \Psi^1(h), 
\end{equation}
where we normalized $\int_0^\infty dh\Psi^{0,1}(h)=1$. Such a variation could reflect the crystalline structure of the solid, which gives rise to a spatial patterns of the solid density $\rho_s$ on molecular scale. The chosen form can be seen as a perturbation expansion for small inhomogeneities or as part of a Fourier decomposition of the chemical variations. 

The liquid-liquid interaction is completely unaffected by this surface inhomogeneity, so we can evaluate the condition ${\mathcal G}=0$, 
\begin{eqnarray}\label{hysteresis}
\cos \theta_\infty &=& \cos \theta_Y + \epsilon  G_q(\alpha),
\end{eqnarray}
where
\begin{eqnarray}
G_q(\alpha) = \int_0^\infty dx \, h'(x)  \Psi^1(h) \,\cos(qx+\alpha).
\end{eqnarray}
It is clear from $(\ref{hysteresis})$ that the macroscopic angle $\theta_\infty$ can take a range of values, and depends sensitively on the phase $\alpha$ at the contact line position (taken at $x=0$ by definition). 

The calculation of $G_q(\alpha)$ requires the slope $h'(x)$ of the nontrivial equilibrium solution. One can, however, get an estimate for the hysteresis by introducing the approximation $h'(x)\approx \tan \theta_Y$:
\begin{eqnarray}\label{galpha}
G_q(\alpha) \approx \int_0^\infty dh \, \Psi^1(h) \,\cos(qh/\tan\theta_Y + \alpha) 
\nonumber \\
= 
\cos \alpha \, \Re\left( \tilde{\Psi} \left(\frac{q}{\tan \theta_Y} \right)\right)
+\sin \alpha \,\Im \left(\tilde{\Psi} \left(\frac{q}{\tan \theta_Y} \right)\right),
\nonumber \\
&& 
\end{eqnarray}
where 
\begin{equation}
\tilde{\Psi}(q) = \int_0^\infty dh \, e^{iqh}\, \Psi^1(h),
\end{equation}
is the Fourier transform of $\Psi^1(h)$ times the Heaviside step function. The amplitude of the hysteresis, characterized by the advancing and receding contact angles $\theta_a$ and $\theta_r$, can then be estimated by 
\begin{eqnarray}
\cos \theta_a &\approx& \cos \theta_Y - \epsilon \left| \tilde{\Psi}\left(\frac{q}{\tan \theta_Y} \right)  \right|, \\
\cos \theta_r &\approx& \cos \theta_Y + \epsilon \left| \tilde{\Psi}\left(\frac{q}{\tan \theta_Y} \right)  \right|.
\end{eqnarray}

This result has two intuitive limits. Since the length scale appearing in $\Psi^1(h)$ will again be $\ell$, the hysteresis is controlled by the dimensionless wavelength 
\begin{equation}
\mathcal{Q}=\frac{q\ell}{\tan \theta_Y}.
\end{equation}
In the limit $\mathcal{Q}\ll 1$, one finds $|\tilde{\Psi}|=1$, because of normalization. It this case the variation of the surface chemistry is slow and one can locally treat the surface as homogeneous, with the relevant surface tensions varying between $\pm \epsilon$. This is the usual 'macroscopic' interpretation of hysteresis based on surface tensions. In the opposite limit, $\mathcal{Q}\gg1$, one finds $|\tilde{\Psi}|\propto 1/\mathcal{Q}\rightarrow 0$. This scaling is due to the singularity in the Heaviside function, which determines the large scale behavior of the Fourier transform. In this limit, the spatial variation is so quick that it is averaged out and the hysteresis disappears altogether. 

\subsection{An illustration}

\begin{figure}[t!]
\includegraphics{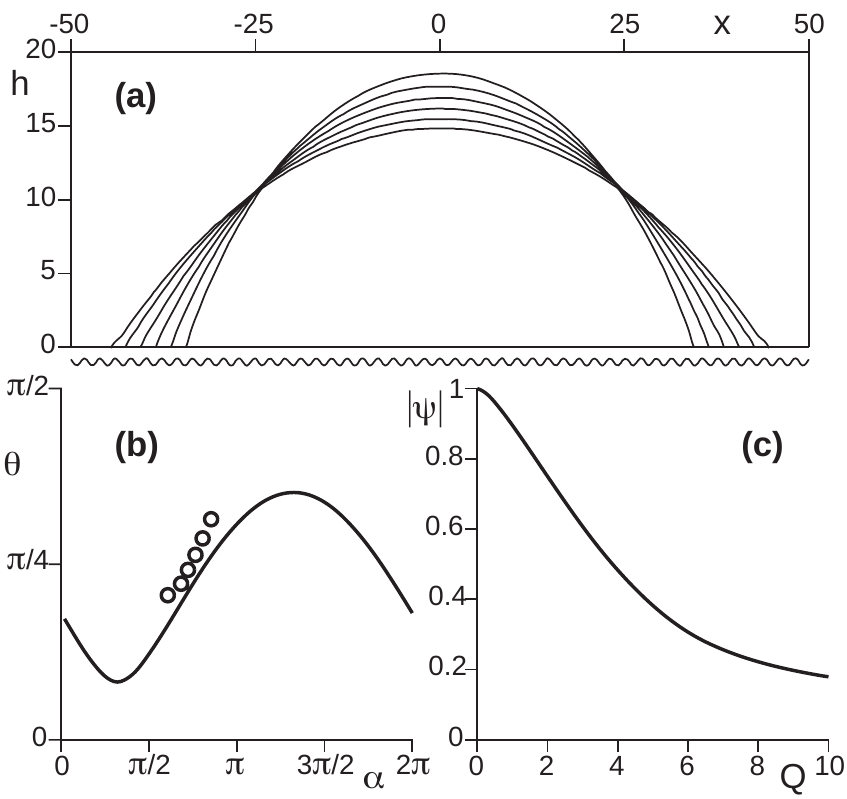}
\caption{(a) Metastable solutions for drops of equal volume in the presence of hysteresis. The phase of the interface potential is sketched below (note that the solid-liquid interface is still taken perfectly flat). Simulation parameters: $\theta_Y=\pi/4$, $\epsilon=0.427$, $\mathcal{Q}=3$. (b) Contact angle versus the phase of the potential $\alpha$. The numerical values (circles) and the approximation (\ref{galpha}). (c) Absolute value of $|\tilde{\Psi}|$, approximating the strenght of the hysteresis for wavelength $\mathcal{Q}$.}  
\label{fig.Hysteresis} 
\end{figure}

The approximation above will now be compared to complete numerical solutions in the presence of hysteresis (Fig.~\ref{fig.Hysteresis}a). We again consider the interaction (\ref{interaction}), for which we can compute $\Psi^0(h)$ for a homogeneous substrate. We consider spatial variations characterized by $\Psi^1(h)=\Psi^0(h)$, which gives (Appendix~\ref{app.numerics}):
\begin{eqnarray}\label{variation}
\Psi^1(h) = \frac{2}{9\ell}
\left\{ 
\begin{array}{l l}
(\ell/h)^3 & \quad \textrm{for $h \geq \ell$} \\
8- 9(h/\ell)+2 (h/\ell)^3 &  \quad \textrm{for $h < \ell$}.
\end{array}
\right.
\end{eqnarray}
This function is indeed normalized to unity. The amplitude of the hysteresis can now be computed by evaluating the integral transform $|\tilde{\Psi}|$, which is plotted in Fig.~\ref{fig.Hysteresis}c as a function of $\mathcal{Q}$. 

We have numerically simulated droplets for this interaction. We have imposed the drop center to coincide with a maximum the wall potential has a maximum, so that mirror symmetry yields two identical contact lines. The volume was kept constant, yielding a discrete set of solutions of varying contact angle $\theta_\infty$. As can be seen from Fig.~\ref{fig.Hysteresis}a, the contact line positions for these metastable solutions are separated by roughly one wavelength. The drops shown in this figure are the only ones we could obtained for this volume, reflecting the bounds of hysteresis. 

As expected from our analysis, the precise phase $\alpha$ of the potential is slightly different for the various drops. This is shown in Fig.~\ref{fig.Hysteresis}b, in which the numerical results are compared to (\ref{galpha}). This approximation accounts quite well for the value of $\theta_a$, as well as for the phase shift with respect to $\Pi_{sl}$, whose extremal values are located at $\alpha=0$ and $\Pi$. However, it significantly underestimates $\theta_r$. We attribute this to the strong deviation from the wedge approximation near the contact line (c.f. the flattest drop in Fig.~\ref{fig.Hysteresis}a).

\section{Discussion}

\subsection{Internal selection of the contact angle}
We have shown how the macroscopic contact angle emerges from a microscopic force balance, within a mean field continuum discription. There is a crucial role for the integral $\mathcal{G}$, which is invariant along the entire interface. Physically, this invariant expresses horizontal force balance on an arbitrary cross-section of the liquid. The macroscopic angle can be computed by equating the value of $\mathcal{G}$ for a macroscopic wedge to its value at the contact line. The height over which the integral attains its asymptotic value is simply $\ell$, the range of the interactions. In this respect, the contact angle is indeed determined at a molecular scale.

A striking feature of the microscopic model is that the entire solution follows from the pressure balance, without the necessity of an external boundary condition. Hence, there is an {\em internal} selection of the contact angle. This is very different from the usual macroscopic theory, for which the contact angle has to be imposed as a boundary condition. Besides this conceptual difference, the internal selection also works when surface tensions are no longer well-defined. This occurs for contact angle hysteresis in the case where variations of the surface chemistry are of the same scale as the range of interaction (expressed in Sec.~\ref{sec.hysteresis} as $\mathcal{Q}\sim 1$). Again using the invariant $\mathcal{G}$, the microscopic model naturally provides a range of contact angles.

At this point we would like to stress that the free energy analyzed in this paper oversimplifies the physics near a contact line: it ignores thermal fluctuations, while the continuum approach starts to break down at molecular scales~\cite{HFC89}. The present analysis should therefore be considered as a model calculation, where one can explicitly identify the selection mechanism. It would be interesting to combine the present approach with molecular dynamics simulations, or more rigorous density functional calculations to improve the physical reality. The results for the (submolecular) angle $\theta_\mu$, which resolved the paradox of the straight wedge solutions, should be interpreted within this context. Having said that, there exist experimental measurements of variations between macroscopic and microscopic angles, for systems with sufficiently long ranged interactions~\cite{PH00,HPF00}. A very similar situation is encountered in electrowetting, where an applied electric field induces a change of the macroscopic contact angle~\cite{L75}. It was recently shown, however, that close to the contact line one recovers Young's angle~\cite{BHM03,MB07}.

\subsection{Local approximations}

We are now in the position to test local approximations based on a free energy functional  $E=\int dx \Gamma(h,h')$. Such a formulation has the great practical advantage that the equilibrium condition reduces to a second order differential equation rather than a nonlocal integral equation. The standard approach based on a disjoining pressure~\cite{DI93,S98} yields $\theta_\mu=0$, independent of $\theta_Y$. We refer to Appendix~\ref{app.local} for details. This is often interpreted in terms of a precursor film, which naturally has $h'=0$, although strictly speaking this is not necessary. Comparing to numerical profiles of the nonlocal theory, we see that this provides a good approximation for small contact angles. 

The approximation clearly fails for large angles, for which the profile is much closer to a straight wedge, and $\theta_\mu \neq 0$. We therefore suggest an alternative local approximation:
\begin{equation}\label{newform}
\Gamma(h,h') = \gamma \left( \sqrt{1+h'^2} -\cos \theta_Y\right)f(h),
\end{equation}
which has perfect wedge solutions, i.e. $\theta_\mu=\theta_Y$. We again refer to Appendix~\ref{app.local} for details. The numerical solutions obtained in Sec.~\ref{sec.numerics} appear to be in between the straight wedge solutions and those from the standard models with disjoining pressure. The approximation for $\theta_\mu$ in (\ref{thetamu1}) has been obtained from interpolation between these two local models. The advantage of the form (\ref{newform}) is that it provides an internal selection of the contact angle. This means that when applied to the nonequilibrium situation of a moving contact line, the solution is self contained and no longer requires an additional condition borrowed from equilibrium. It remains to be investigated how the local approximations compare to the nonlocal theory in the dynamical case.

Let us conclude by noting that we have not been able to come up with a self-consistent derivation of a local approximation of the free energy for arbitrary $\theta$. As suggested in~\cite{GD98}, this is perhaps not possible due to the intrinsically nonlocal character of the problem. 

{\bf Acknowledgements\\}
The authors are indebted to Jean-Fran\c{c}ois Joanny for pointing out the importance of the invariant $\mathcal{G}$ for the selection of the contact angle. We gratefully acknowledge continued support by Jens Eggers and Philippe Brunet, as well as discussions with Bob Evans, Julien Tailleur and Valeriy Slastikov. B.A. thanks Jean-Louis Barrat and Bernard Castaing to whom the puzzle of Hocking's exact solutions violating Young's law was presented. J.H.S. acknowledges financial support by a Marie Curie European Grant FP6 (MEIF-CT2006-025104).

\begin{appendix}

\section{Calculation of $\mathcal{G}_{ll}$}\label{app.MK}

\begin{figure}[t!]
\includegraphics{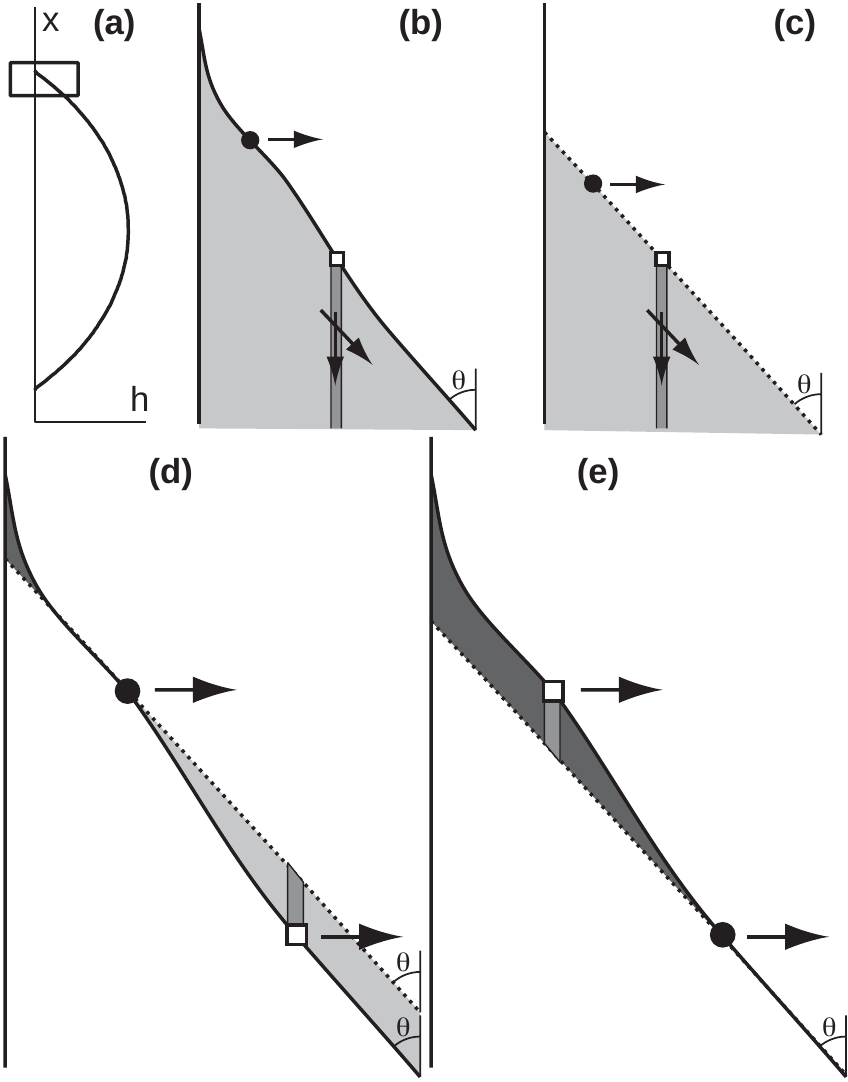}
\caption{\textbf{(a)} $\mathcal{G}_{ll}$ can be interpreted as the total force on a liquid corner due to liquid-liquid interactions. We define $h$ as the distance to the solid surface. \textbf{(b)} Graphical representation of $\mathcal{G}_{ll}$ as the potential at the interface at $z=h$, (black dot), integrated over all $h$ (moving the black dots along the interface). The potential is decomposed as the contribution of the dark grey rectangular zone, summed over all positions of the square. \textbf{(c)} We compare the result for (b) to that for a liquid wedge of angle $\theta$. \textbf{(d)} The differce beteen (b) and (c) is expressed by integration over a new domain (see text), consisting of positive (light gray) and negative (dark grey) contributions. \textbf{(e)} Inverting the two moving points, the contribution is identical but with opposite signs.}
\label{fig.MerchantKeller} 
\end{figure}

We provide a graphical representation of the analysis by Merchant and Keller~\cite{MK92} to compute $\mathcal{G}_{ll}$. Expression (\ref{gll}) can be seen as the integral over $h$, represented schematically by the black dot moving along the interface (Fig.~\ref{fig.MerchantKeller}b), of the potential energy at the interface evaluated at the position of the black dot. This potential itself is an integral over the entire liquid volume (light gray zone) of the intermolecular potential. Note that by considering $\varphi_{ll}$ we already integrated out the invariant direction parallel to the contact line. Diagramatically, we have represented the potential as the interaction of the black dot with the dark grey rectangle, integrated over all positions of the open square. 

The trick is to compare the result for an arbitrary shape that tends to a given angle $\theta$ (Fig.~\ref{fig.MerchantKeller}b), to that for a perfect liquid wedge of same angle (Fig.~\ref{fig.MerchantKeller}c). The differnce between the two cases can be expressed as an integral over $h$ (moving black dot) over the domain sketched in 
Fig.~\ref{fig.MerchantKeller}d. This domain extends from the real interface (solid line) to the plane of angle $\theta$ (dotted line). The difference is counted positively if the interface is below the plane (light gray) and negatively otherwise (dark gray). If we interchange the position of the solid dot and the square dot, the potential energy due to the rectangle is exactly the same, but with opposite sign (Fig.~\ref{fig.MerchantKeller}e). Since both contributions are encountered through the double integration, the total difference vanishes by antisymmetry. This means that we can directly compute $\mathcal{G}_{ll}$ by considering the simple wedge geometry, which gives the result (\ref{gllbis})~\cite{MK92}. Note that this argument requires the assumption that the interaction depends only on the distance between two points, but not on the spatial location of the points. If the interaction were inhomogeneous, the perfect antisymmetry would be broken.

\section{Local approximations}\label{app.local}

The usual local approximation is to add the Laplace pressure and the disjoining pressure, giving the equilibrium condition
\begin{equation}
-\gamma \kappa + \Pi_{\rm disj}(h) = \lambda.
\end{equation}
This can be derived from the free energy 
\begin{equation}
\Gamma_1(h,h') = \gamma\left( \sqrt{1+h'^2} -1 \right) + \gamma (1-\cos \theta_Y) f(h),
\end{equation}
through the standard Euler-Lagrange description (see Sec.~\ref{sec.macro}). The function $f(h)$ is the integrated disjoining pressure, renormalized such that $f(0)=0$, $f(\infty)=1$. The invariant corresponding to this free energy reads
\begin{equation}
{\mathcal G}_1 = \gamma \left( \frac{1}{(1+h'^2)^{1/2}} -1 \right) + \gamma (1-\cos \theta_Y) f(h),
\end{equation}
so that the equilibrium condition ${\mathcal G}_1=0$ yields $\theta_\mu=0$, independent of $\theta_Y$. 

For larger angles we suggest an alternative local approximation:
\begin{equation}
\Gamma_2(h,h') = \gamma \left( \sqrt{1+h'^2} -\cos \theta_Y\right)f(h),
\end{equation}
which has a capillary pressure
\begin{equation}
\Pi_2 = -\gamma \kappa f(h) - \frac{df}{dh}\left( \frac{1}{(1+h'^2)^{1/2}} -\cos \theta_Y \right)
\end{equation}
and a first integral
\begin{equation}
{\mathcal G}_2 =  \gamma \left( \frac{1}{(1+h'^2)^{1/2}} -\cos \theta_Y \right) f(h).
\end{equation}
The equilibrium solution has $\theta=\theta_Y$ for any value of $h$, indeed representing a straight wedge. In particular $\theta_\mu=\theta_Y$.

The solution of the nonlocal equation appears to be bounded by the local approximations obtained from $\Gamma_1$ and $\Gamma_2$. To estimate $\theta_\mu$, we therefore propose a linear interpolation 
\begin{equation}\label{interpolate}
\theta_\mu = \nu \theta_{\mu,1} + (1-\nu)\theta_{\mu,2}=(1-\nu)\theta_Y.
\end{equation} 
The weight $\nu$ can be estimated from comparison to the pressure on a straight wedge of angle $\theta_Y$, Eq.~(\ref{piwedge}), as computed by Hocking~\cite{H93}. Even though Hocking's wedge is not an equilibrium solution, it does provide the correct nonlocal asymptotic disjoining pressure for $h\gg \ell$. For the second local approximation one finds $\Pi_2=0$ for a wedge of $\theta_Y$, while the first approximation asymptotically gives (\ref{piwedge}) without the contribution $F(\theta)$. We therefore equate
\begin{equation}
\nu \left( 1 -   \frac{c_{sl}}{c_{ll}}      \right) = 
\left( 1 -   \frac{c_{sl}}{c_{ll}} - F(\theta)     \right).
\end{equation}
Combined with (\ref{interpolate}), this gives prediction (\ref{thetamu1}) shown in Fig.~\ref{fig.ThetaMicro}.

\section{Numerical solution of drop profiles}\label{app.numerics}

This appendix provides details on the numerical resolution of equilibrium drop shapes. We consider the effective interaction 
\begin{eqnarray}
\tilde{\phi}_{\alpha \beta}(r) =
\left\{ 
\begin{array}{l l}
-c_{\alpha \beta}/r^6 & \quad \textrm{for $r \geq \ell$} \\
-c_{\alpha \beta}/\ell^6 &  \quad \textrm{for $r < \ell$}.
\end{array}
\right.
\end{eqnarray}
The constants $c_{sl}$ and $c_{ll}$ can be related to the surface tensions, using (\ref{spreading}):
\begin{eqnarray}
\gamma &=& \frac{3\pi c_{ll} }{8 \ell^2} \\
\gamma(1+\cos\theta_Y) &=& \frac{3\pi c_{sl} }{4 \ell^2}.
\end{eqnarray}

\subsection{Algorithm}

We perform the following iterative procedure for $h(x)$ in order to find the equipotential surface. At step $n$, the (discretized) shape of the drop is noted $h_n(x)$. The potential corresponding to this shape, $\Pi_n(x)$, (see the second part of the appendix). The shape is evolved according to the equation:
\begin{equation}
h_{n+1}(x)=h_{n}(x)+\delta \left(<\Pi_n> -\Pi_n(x)\right)
\end{equation}
where $<\Pi_n>$ is the average potential. The parameter $\delta$ is fixed at a value sufficiently small to ensure numerical stability. The positions of the contact lines are evolved to ensure the mass conservation between steps: the drop spreads when the potential at the contact line is smaller than its average value; otherwise, it shrinks.

In practice, we have observed that the result becomes independent of the discretisation when the mesh size becomes lower than $0.05$ ($=0.03$ for the figures shown here).  The convergence of the calculation, starting from a spherical cap takes approximately $300$ steps ($1000$ for the figures shown here). The spatial variations of the potential then reach the numerical noise level.

\subsection{Evaluation of the pressure}

We evaluate the pressure according to (\ref{pcap}). The second integral is constant and can thus be ignored. As the effective interaction is defined over two domains, we need to to separate zones of integration into those that have $r \geq \ell$ and $r< \ell$ respectively. First consider the solid-liquid interaction, i.e. the third integral of (\ref{pcap}). If $h\geq \ell$, the integral is easily performed as 
\begin{eqnarray}
\Pi_{sl}(h)&=&\int_{0}^{\pi/2}d\theta\int_{0}^{2\pi}d\varphi\int_{h/\cos\theta}^{+\infty}  r^2 \sin \theta \tilde{\phi}_{sl}(r) dr 
\nonumber \\
&=&-\frac{\pi c_{sl}}{6 h^3}.
\end{eqnarray}
For $h<\ell$, the domain has to be separated into various regions. Introducing $\cos\alpha=h/\ell$, we get for $h< \ell$:
\begin{eqnarray}
\Pi_{sl}(h)= \nonumber \\
-2\pi c_{sl} \int_{\alpha}^{\pi/2} d\theta \sin \theta\int_{h/\cos\theta}^{+\infty}  dr/r^4\nonumber\\
-2\pi c_{sl} \int_{0}^\alpha d\theta \sin \theta \left(\int_{h/\cos\theta}^{\ell}  r^2dr/\ell^6+\int_{\ell}^{+\infty}  dr/r^4\right),\nonumber \\
\end{eqnarray}
which reduces to (for $h<\ell$),
\begin{equation}
\Pi_{sl}(h)=-\frac{\pi c_{sl}}{6\ell^3} \left(8- 9(h/\ell)+2 (h/\ell)^3\right).
\end{equation}
This potential $\Pi_{sl}$ is indeed continuous at $h=\ell$, and it obeys the property:
\begin{equation}
\int_0^\infty dh \Pi_{sl}(h)=\frac{3\pi c_{sl}}{4\ell^2}=\gamma_{sl}-\gamma-\gamma_{sv}=-\gamma (1+\cos \theta_Y)
\end{equation}
\begin{figure}[t!]
\includegraphics{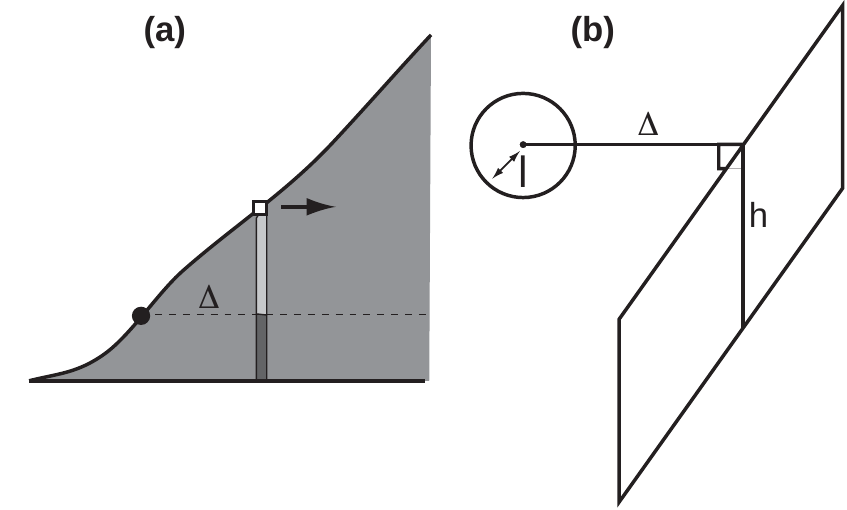}
\caption{(a) The interaction potential at the black point, due to the liquid zone at the same $x$ as the open square (rectangular zone), is decomposed into the two regions shown in light and dark gray. (b) The elementary configuration allowing to express the potential due to the liquid-liquid interaction is its contribution due to a band of liquid of height $h$ at a distance $\Delta$ from the point considered.} 
\label{fig.SphericalCap} 
\end{figure}

Provided that we realise that $\gamma_{ll}=0$, the same formula can be applied to the case of the liquid-liquid and solid-solid interactions:
\begin{equation}
c_{ll}=\frac{8 \gamma \ell^2}{3 \pi}
\end{equation}
The liquid-liquid potential can be expressed as:
\begin{equation}
\Pi_{ll}=\int_{a}^{b}dx' \int_{-\infty}^{\infty}dy'\int_{0}^{h(x')}  \tilde{\phi}_{ll}(r)dz'
\end{equation}
with $r^2=(x'-x)^2+y'^2+(z'-h(x))^2$. The integral over $z$ can be decomposed into two contributions of the same form, as shown in Fig.~\ref{fig.SphericalCap}). Introducing the energy of interaction $\zeta(\Delta,h)$ between a point and a rectangular zone of height $h$, infinite in the transverse direction, at a distance $\Delta$, 
\begin{equation}
\zeta(\Delta,h)= \int_{-\infty}^{\infty}dy'\int_{0}^{h)}  \tilde{\phi}_{ll}\left(\sqrt{\Delta^2+y'^2+z'^2}\right)dz'
\end{equation}
we get:
\begin{equation}
\Pi_{ll}(x)=\int_{-\infty}^\infty d\Delta \left(\zeta(\Delta,h(x))+\zeta(\Delta,h(x+\Delta)-h(x))\right)
\end{equation}
Importantly, $\zeta$ is defined as an odd function of $h$ ($\zeta(\Delta,-h)=-\zeta(\Delta,-h)$) but an even function of $\Delta$. In the situation where $\Delta>\ell$, we introduce $\cos \alpha=\left(1+\frac{h^2}{\Delta^2 \cos^2 \varphi}\right)^{-1/2}$ and get:
\begin{equation}
\zeta=-\frac{16 \gamma \ell^2}{3\pi} \int_{0}^{\pi/2} d\varphi \int_{0}^\alpha \frac{\cos^6 \theta}{\Delta^6} \Delta^2 \frac{\sin \theta}{\cos^3 \theta}d\theta \nonumber
\end{equation}
It simplifies into:
\begin{equation}
\frac{\zeta}{\gamma}=- \frac{2 h^2+3 \Delta^2}{3 \Delta^4 \left(h^2+\Delta^2\right)^{3/2}} \ell^2 h \;\; \mbox{for} \;\; \Delta>\ell
\end{equation}
Now for $\Delta<\ell$, there is a circular screening zone of radius $\sqrt{\ell^2-\Delta^2}$. First situation, the screening zone is smaller than the height:
\begin{equation}
\zeta=-\frac{16 \gamma \ell^2}{3\pi} \int_{0}^{\pi/2} d\varphi \left[ \int_{\cos \alpha}^{\Delta/\ell} \frac{x^6}{\Delta^6} \frac{\Delta^2  dx}{x^3}+ \int_{\Delta/\ell}^1 \frac{1}{\ell^6} \frac{\Delta^2  dx}{x^3} \right] \nonumber
\end{equation}
For $\ell^2>\Delta^2>\ell^2-h^2$, it simplies into:
\begin{equation}
\frac{\zeta}{\gamma}=- \frac{(2 h^2+3 \Delta^2) \ell^2 h}{3 \Delta^4 \left(h^2+\Delta^2\right)^{3/2}}+\frac{2\ell^2}{3\Delta^4} + \frac{4 \Delta^2}{3 \ell^4}-\frac{2}{\ell^2}
\end{equation}
Second situation, the screening zone radius $\sqrt{\ell^2-\Delta^2}$ is larger than the height $h$. We define $\cos \varphi_0=h/\sqrt{\ell^2-\Delta^2}$ and get:
\begin{eqnarray}
\zeta&=&-\frac{16 \gamma \ell^2}{3\pi}\left[ \int_{0}^{\varphi_0} d\varphi  \int_{\cos \alpha}^1 \frac{1}{\ell^6} \frac{\Delta^2  dx}{x^3} \right. \nonumber\\
&+&\left . \int_{\varphi_0}^{\pi/2} d\varphi \left( \int_{\cos \alpha}^{\Delta/a} \frac{x^6}{\Delta^6} \frac{\Delta^2  dx}{x^3}+ \int_{\Delta/a}^1 \frac{1}{\ell^6} \frac{\Delta^2  dx}{x^3} \right) \right] \nonumber
 \end{eqnarray}
 For $\ell^2-h^2>\Delta^2$, It finally simplies into:
\begin{eqnarray}
\frac{\zeta}{\gamma}&=& \frac{2 h \sqrt{\ell^2-\Delta^2-h^2} }{3  \left(\Delta^2+h^2\right) \pi } \left(\frac{1}{ \Delta^2}- \frac{4 \left(\Delta^2+h^2\right)}{\ell^4}\right)\nonumber\\
&-&\frac{2 \ell^2 h \left(3 \Delta^2+2 h^2\right) \sin^{-1}\left(\sqrt{\Delta^2+h^2}\right)}{3 \Delta^4 \left(\Delta^2+h^2\right)^{3/2} \pi }\\
&+&\frac{4}{3  \pi } \left(\frac{\ell^2}{ \Delta^4}-\frac{3}{\ell^2}+\frac{2 \Delta^2}{\ell^4}\right) \sin^{-1}\left(\frac{h}{\sqrt{\ell^2-\Delta^2}}\right) \nonumber
 \end{eqnarray}

\end{appendix}


\begin{thebibliography}{99}
\bibitem{footMK} The interactions $\tilde{\phi}_{ll}$ and $\tilde{\phi}_{sl}$ considered in~\cite{MK92} were of identical form, had identical regularization lengths, $\ell_{ll}=\ell_{sl}$, but different amplitudes $c_{ll}\neq c_{sl}$. In this situation one can thus apply (\ref{youngmicro}) to compute Young's angle. From the caption it seems that Fig.~3a of~\cite{MK92} uses $c_{sl}/c_{ll}=0.9944$ with an asymptotic boundary condition $\theta_\infty=135^\circ$. Comparing to (\ref{youngmicro}) this angle is not $\theta_Y$.
\end{thebibliography}
\end{document}